# Null expectations and null hypothesis testing for the species abundance distribution

Gabriel Arellano
gabriel.arellano.torres@gmail.com
Ecology and Evolutionary Biology, University of Michigan, Ann Arbor, Michigan 48109, USA
Version: 03-March-2021

**Abstract** · The number of elements ($N$) and types ($S$) sampled from an ecological system are among the most powerful constraints on observations of abundance, distribution, and diversity. Together, $N$ and $S$ determine sets of possible forms (i.e., feasible sets) for the species abundance distribution (SAD). There are three approaches to the description of the null SAD (= the average feasible SAD). The first approach is based on the random uniform sampling of surjections. I calculate the probability of a given SAD, given $N$ and $S$, under this approach (Eq. 4). The second approach is based on the random sampling of compositions. I calculate the probability of a given SAD, given $N$ and $S$, under this approach (Eq. 8). The third approach is based on the random uniform sampling of partitions. I review the approach, which was developed by Locey & White (2013), and provide some asymptotic results useful for ecologists. The center of a feasible set is a null expectation, which should deviate enough from the alternative models before invoking for biological or ecological mechanisms underlying the SAD. Here, I integrate the feasible set approach with the typical framework of inference in ecology (goodness-of-fit, null hypothesis testing, model comparison, null modelling). I describe how to perform numerical simulations to describe expectations under different approaches to the feasible set. I develop objective or fitness functions to allow the estimation of the most likely SAD using numerical optimization. I provide tools to compare null expectations based on the feasible set approach with the observations, in the context of model comparison and null hypothesis testing.

## Introduction

The species abundance distribution (SAD) is a fundamental descriptor of any biological community. It can be defined as the vector of comparable abundances for the species in a community (McGill 2011). It is almost a universal law that most species are rare and few are common (the 'hollow curve' SAD). Many different theoretical models compete to explain the SAD in biological communities. Most of them appeal to biological or ecological mechanisms (McGill *et al.* 2007). However, the existence of hollow curves in non-biological systems (e.g. Nekola & Brown 2007; Warren *et al.* 2011) suggests that the overall shape of natural SADs results from statistical phenomena alone. There are several purely statistical models that rely on various assumptions, other than biological mechanisms (McGill 2003a, 2010; Šizling *et al.* 2009; White, Thibault & Xiao 2012; Conlisk *et al.* 2012). Perhaps the most popular is the log-normal SAD, invoking to random multiplicative process acting on abundances (Harte 2003; McGill 2003b; Nee & Stone 2003). The present work builds on Locey & White (2013), who provided a simpler and more intuitive description of the expected form of the SAD based solely on $N$ and $S$ and without assuming any ecological or statistical process, whether random or non-random. From their perspective, the expected SAD is just the "typical SAD", among all the possible SADs.

When sampling species abundances, usually we have a sample with $S$ species and $N$ individuals, each species $i$ with abundance $n_i \geq 1$, so that $\sum_{i=1}^{S} n_i = N$. The SAD of such community is the vector $\{n_1, n_2, n_3, \ldots, n_S\}$. There are many possible forms of the sample SAD for a given $N$ and $S$. For a sample of $S = 5$ and $N = 15$, one possible SAD is $\{3, 3, 3, 3, 3\}$, other is $\{6, 4, 3, 1, 1\}$, other is $\{5, 4, 3, 2, 1\}$, etc. Likewise, there are many possible ways to obtain a specific form of the SAD, e.g., $\{6, 4, 3, 1, 1\}$ yields the same frequency distribution and the same rank-abundance curve as $\{1, 3, 1, 6, 4\}$. The set of all possible SADs is called the feasible set (Locey & White 2013). If the feasible set is small, then there is little information in the SAD being examined, and it will be difficult or impossible to determine the processes that generated it (Haegeman & Loreau 2008; Locey & White 2013). Not only the size of the feasible set is important, but also the distribution of the shapes within it. If theoretical predictions do not deviate significantly from the most common shapes in the feasible set (the "center" of the feasible set), then they provide little information about processes (Locey & White 2013).

Comparing the expectations by more complex ecological or statistical models with the expectations by $N$ and $S$ alone is a necessary step before attributing any given observed SAD shape to a given statistical or ecological mechanism. Although this approach is highly promising, and able to generate parsimonious explanations of empirical datasets, most relevant results are scattered across texts on combinatorics, gray literature, or rather obscure mathematical ecology related to entropy maximization techniques. Here, I compile and digest combinatorial results for the ecological audience and develop new results when necessary. I also present tools to integrate the feasible set approach with the typical framework of inference in ecology (goodness-of-fit, null hypothesis testing, model comparison, null modelling). See Table 1 for a quick summary.

## Integer surjections, compositions, and partitions

We can imagine any sampling effort as a random draw of one SAD from the set of all possible SADs having the same $N$ and $S$, in the same way as a person can draw one ball at random from an urn with many balls. If our sample is compatible with a random draw from an urn containing all possible SADs, then we cannot attribute the form of the pattern to any biological or ecological mechanism without additional evidence. There are at least three ways of defining the content of the urn, which have different interpretations and link to three different null hypotheses.

First, the urn may contain sequences of individuals (not species), with the taxonomic identity of each individual. If $N = 5$, there will be one ball for $(A, A, B, B, B)$, another different ball for $(A, B, A, B, B)$, another different ball for $(A, B, B, A, B)$, and so on, being $A$ and $B$ two different species. This type of "balls" are called *integer surjections*.

Second, the urn may contain sets of species-level abundances with respect to which species has which abundance. Using the same example as above, there will be one ball for $\{A = 2, B = 3\}$ and another different ball for $\{A = 3, B = 2\}$. But there will be just one ball ($\{A = 2, B = 3\}$) to represent all the possible integer surjections ($(A, A, B, B, B)$, $(A, B, A, B, B)$, $(A, B, B, A, B)$, etc). This type of "balls" are called *integer compositions*.

Third, the urn may contain sets of species-level abundances without respect to which abundance belongs to which species. That is, there will be one ball for $\{3, 2\}$ and a different ball for $\{4, 1\}$. But there will be just one ball ($\{3, 2\}$) to represent all the possible integer compositions ($\{A = 2, B = 3\}$ and $\{A = 3, B = 2\}$) and the corresponding integer surjections. This type of "balls" are called *integer partitions*.

A microstate is one of the ways in which a macrostate can be expressed. So defined, integer surjections are microstates of integer compositions, and integer compositions are microstates of integer partitions. Integer partitions are the highest level of SAD macrostate. In the case of SAD studies, ecologists always focus on the partition corresponding to the observed SAD. Labels are removed and the species are sorted from the most to the least abundant (McGill *et al.* 2007; McGill 2011), so individual and species names does not matter. Dropping information from the surjection and composition levels to the partition level is necessary to make the SAD useful to compare different communities. Because this is common practice in the study of SADs, one needs to calculate always the probability of integer partitions, even when sampling surjections or compositions.

The following sections describe the three null hypotheses that result from the three different definitions of "ball in the urn": $H_{null}^{surj}$, $H_{null}^{comp}$ and $H_{null}^{part}$. The selection of the appropriate null hypothesis cannot be done just based on a priori considerations, and should be done by comparing with the empirical data (Haegeman & Etienne 2010). $H_{null}^{surj}$ predicts very even SADs (tending to completely even as $N \to \infty$), $H_{null}^{comp}$ predicts reasonably uneven SAD shapes (tending to the classical broken stick model as $N \to \infty$), and $H_{null}^{part}$ predicts very uneven SADs, similar to the log-normal model and empirical observations (Locey & White 2013). Fig. 1 contains a quick comparison between them.

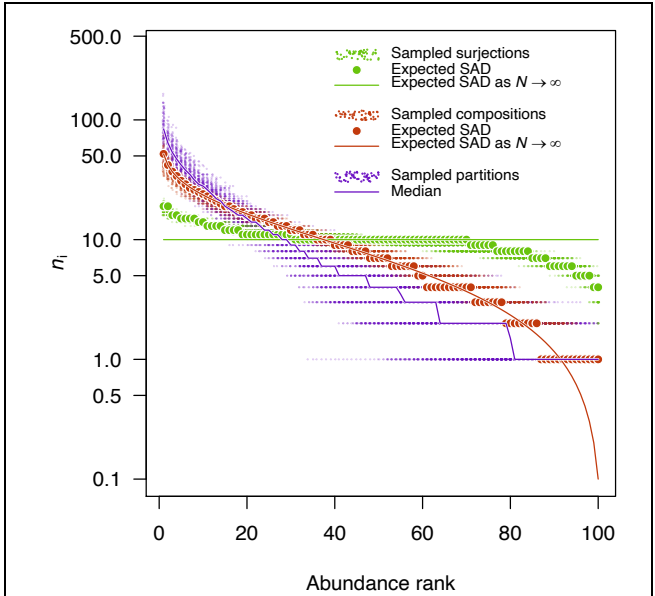

**Figure 1.** Comparison between the expectations by the uniform random sampling of integer surjections ($H_{null}^{surj}$), the uniform random sampling of integer compositions ($H_{null}^{comp}$), and the uniform random sampling of integer partitions ($H_{null}^{part}$). The expectations are based on $N = 1000$ individuals and $S = 100$ species. The small dots are direct samples from the feasible sets. They have been jittered a bit for visualization purposes. The big dots represent the most likely SADs after numerical optimization (see Appendix 3 for the code used). Note that there is not such a thing as "the most likely SAD" under $H_{null}^{part}$. Lines represent asymptotical solutions (for $H_{null}^{surj}$ and $H_{null}^{comp}$) or the median abundance of the $i^{th}$ most common species when sampling partitions under $H_{null}^{part}$.

**First null hypothesis: sampling surjections, $H_{null}^{surj}$**

*Probability of a given SAD*

Surjections are the result of throwing $N$ times a $S$ sided dice, as long as each of the species appears at least once. The probability that each of the $N$ individuals will receive any given species label is $1/S$, since all the species are equally likely. The probability of a given composition from the random picking of surjections is given by the multinomial distribution. This is not enough to compare any given sample with the expected by $H_{null}^{surj}$. We need to define the probability of a given partition (not composition) resulting from the random picking of surjections. In general, the probability of observing a given SAD in the sample, given $N$ and $S$, i:

Eq. 1

$$P(partition\ SAD|N,S) = \frac{number\ of\ surjections\ for\ that\ partition}{number\ of\ surjections}$$

All the compositions resulting in a given partition are associated to the same number of integer surjections, because there is a one-to-one equivalence between the different labels. It does not matter if the composition is $\{n_A = 2, n_B = 3\}$ or $\{n_A = 3, n_B = 2\}$: the number of different words that can be built using the letters AABBB is the same as the number of different words that can be built using the letters BBAAA. Besides, there is no overlap in the sets of surjections that are associated to two different compositions: one cannot write twice the same word unless one has the same letters and in the same frequencies. By definition, different compositions of a partition have the same frequencies, but of different letters. Therefore, we can calculate the number of surjections for a given partition by multiplying the number of compositions per partition by the number of surjections for any composition resulting in that partition.

The number of surjections of any given composition is:

$$\binom{N}{n_1}\binom{N-n_1}{n_2}\binom{N-n_1-n_2}{n_3}\cdots\binom{N-n_1-\cdots-n_{S-1}}{n_S}$$
$$= \prod_{i=1}^{S}\binom{\Sigma_{j=i}^{S} n_j}{n_i}$$

Eq. 2

The number of compositions per partition is $\frac{S!}{\prod_{n=1}^{N}\Phi_n!}$, being $\Phi_n$ the number of species with abundance $n$ in the sample (Etienne & Olff 2005).

Combining both expressions gives the

number of surjections that result in the same partition $SAD = \{n_1, n_2, n_3, \ldots, n_S\}$, $n_1 \geq n_2 \geq n_3 \geq \ldots \geq n_S \geq 1$, $\sum_{i=1}^{S} n_i = N$:

Eq. 3
$$\frac{S!}{\prod_{n=1}^{N} \Phi_n!} \prod_{i=1}^{S} \binom{\Sigma_{j=i}^{S} n_j}{n_i}$$

On the other hand, the total number of surjections, $\sigma(N, S)$, is $S! \begin{Bmatrix} N \\ S \end{Bmatrix}$, where $\begin{Bmatrix} N \\ S \end{Bmatrix}$ are Stirling numbers of the second kind. Therefore:

$$P(partition\ SAD|N, S) =$$

Eq. 4
$$\frac{S! \prod_{i=1}^{S} \binom{\Sigma_{j=i}^{S} n_j}{n_i}}{\sigma(N,S) \prod_{n=1}^{N} \Phi_n!} = \frac{\prod_{i=1}^{S} \binom{\Sigma_{j=i}^{S} n_j}{n_i}}{\begin{Bmatrix} N \\ S \end{Bmatrix} \prod_{n=1}^{N} \Phi_n!}$$

*What are the typical SADs expected under $H_{null}^{surj}$? And the most likely SAD?*

I do not know of a closed expression for the most likely partition SAD expected by $H_{null}^{surj}$. However, we can sample randomly the feasible set many times to estimate the distribution of expected SADs under $H_{null}^{surj}$, including the most likely shapes. Under this null hypothesis, the expectations for the SAD are generated by sampling with replacement from $\{1, 2, 3, \ldots, S\}$, and then counting how many times we find the most frequent number, the second most frequent number, and so on. The R code in Appendix 1 generates one null partition SAD, based on the unbiased sampling of one surjection SAD, tracing the levels surjection → composition → partition.

Uniform sampling from the feasible set is a general and intuitive solution, and it is recommended as long as it is computationally feasible. Alternatively, we could estimate the most likely SAD by optimizing $S$ parameters in a way that $P(partition\ SAD|N, S)$ is maximized. For such optimization, it is not necessary to deal with the huge numbers involved in the expression above; at least not with $\begin{Bmatrix} N \\ S \end{Bmatrix}$, which does not depend in the particular abundances of the species. For the optimization, we would need to maximize the following expression:

Eq. 5
$$\log \frac{\prod_{i=1}^{S} \binom{\Sigma_{j=i}^{S} n_j}{n_i}}{\prod_{n=1}^{N} \Phi_n!} = \sum_{i=1}^{S} \log \binom{\Sigma_{j=i}^{S} n_j}{n_i} - \sum_{n=1}^{N} \log \Phi_n!$$

Using the Stirling's approximation to the factorial ($\log n! \approx n \log n - n$) we know that $\log \binom{n}{m} = \log \frac{n!}{m!(n-m)!} = \log n! - \log m! - \log(n-m)! \approx (n \log n - n) - (m \log m - m) - ((n-m) \log(n-m) - (n-m)) = n \log n - m \log m - (n-m) \log(n-m)$. (4) This approximation is undefined when $n = m$, but $\log \binom{n}{m} = 0$ in that case, so we can safely substitute $\sum_{i=1}^{S} \log \binom{\Sigma_{j=i}^{S} n_j}{n_i}$ by $\sum_{i=1}^{S-1} \log \binom{\Sigma_{j=i}^{S} n_j}{n_i}$ in Eq. 5. Then, to estimate the most likely SAD under $H_{null}^{surj}$ we would need to maximize: (5)

Eq. 6
$$\sum_{i=1}^{S-1} \left( \left( \sum_{j=i}^{S} n_j \right) \log \left( \sum_{j=i}^{S} n_j \right) \right)$$
$$- \sum_{i=1}^{S-1} (n_i \log n_i)$$
$$- \sum_{i=1}^{S-1} \left( \left( \sum_{j=i}^{S} n_j - n_i \right) \log \left( \sum_{j=i}^{S} n_j - n_i \right) \right)$$
$$- \sum_{n=1}^{N} (\Phi_n \log \Phi_n - \Phi_n)$$

Eq. 6 involves much smaller terms than Eq. 4 or Eq. 5, and is a rather manageable problem for moderate numbers of individuals and species. Appendix 3 includes a routine using numerical optimization to generate the most likely SAD predicted by $H_{null}^{surj}$ by maximizing Eq. 6. As $N \to \infty$, the expected abundances of the different species are more and more even, and $n_i \to N/S$ for all $i$.

$H_{null}^{surj}$ has limited application in ecology, because it results in very even abundances, very far from most observed SAD shapes. I present it here for three reasons. First, for the sake of completeness. Second, $H_{null}^{surj}$ may be relevant in the study of abundances distributions across different levels of hierarchical structures, if it applies to higher levels or organization (like taxonomic families, functional groups, etc.). Finally, the null hypotheses discussed here ($H_{null}^{surj}$, $H_{null}^{comp}$, $H_{null}^{part}$) are not intended to describe the reality as it is, but as a reference to give meaning to the empirical observations. Different parts of the SAD may deviate from the different feasible sets in different ways (see section "*Comparing the null expectations with observations and alternative models*"). From that point of view, a null hypothesis that is very far from the whole SAD may be useful to interpret or describe patterns in certain parts of the SAD, like the tail of rare species, for example.

## Second null hypothesis: sampling compositions, $H_{null}^{comp}$

### Probability of a given SAD

If we assume that the urn contains composition-SADs, $\{A = 2, B = 3\}$ and $\{A = 3, B = 2\}$ are different cases. But, unlike under $H_{null}^{surj}$, both cases are represented by exactly one ball each. To compare any given sample with the expected by $H_{null}^{comp}$, we need to define the partitions resulting from the random picking of compositions. In this scenario, the probability of a given partition is:

$$P(partition\ SAD|N,S) = \frac{number\ of\ compositions\ for\ that\ partition}{number\ of\ compositions}$$

Eq. 7

As indicated above, the number of compositions for a given partition is $\frac{S!}{\prod_{n=1}^{N} \Phi_n!}$. The number of compositions of $N$ into $S$ positive parts is $\binom{N-1}{S-1}$. Therefore:

$$P(partition\ SAD|N,S) = \frac{\frac{S!}{\prod_{n=1}^{N} \Phi_n!}}{\binom{N-1}{S-1}} = \frac{S!(N-S)!(S-1)!}{(N-1)!\prod_{n=1}^{N} \Phi_n!}$$

Eq. 8

The expression above describes the discrete broken stick model (Etienne & Olff 2005). The mathematical equivalence between the broken stick model and the sampling process from the composition feasible set has been known for long, and is discussed frequently in the entropy maximization literature (Pueyo *et al.* 2007; Haegeman & Loreau 2008; Locey & White 2013). As suggested by MacArthur himself (1957), deviations from the continuous broken stick model would inform about ecological or biological processes, like the presence of more than one assemblage in the sample.

### *What are the typical SADs expected under $H_{null}^{comp}$? And the most likely SAD?*

The mathematical equivalence between the broken stick model and $H_{null}^{comp}$ greatly facilitates numerical simulations, if we are interested in sampling from the feasible set to estimate the most likely SAD shape and the variability around it. We can simply break the sequence $\{1, 2, 3, ..., S\}$ at random to obtain unbiased random samples from the feasible set of compositions. Appendix 2 contains R code that generates one null partition SAD, based on the unbiased sampling of one composition SAD.

Although uniform sampling from the feasible set is intuitive and recommended, we can also estimate the most likely SAD by optimizing $S$ parameters in a way that $P(partition\ SAD|N,S)$ is maximized. For such optimization, it is not necessary to deal with all the huge numbers involved in Eq. 8. We can minimize $\prod_{n=1}^{N} \Phi_n!$ (the only component there that involves individual abundances) or, equivalently, $\sum_{n=1}^{N} \log \Phi_n! \approx \sum_{n=1}^{N} (\Phi_n \log \Phi_n - \Phi_n)$, using the Stirling's approximation to the factorial. Appendix 3 includes a routine using numerical optimization to generate the most likely SAD predicted by $H_{null}^{comp}$.

In any case, the asymptotic behavior when $N \rightarrow \infty$ is tractable, both while doing numerical simulations (Appendix 2) and analytically. Under $H_{null}^{comp}$, the species rank-abundance distribution predicted by MacArthur's (1957) broken stick model is the most likely partition SAD (Pueyo *et al.* 2007). Using this model, the expected abundance of the $i^{th}$ most abundant species in an assemblage with $S$ species is:

$$n_i = \frac{N}{S} \sum_{x=i}^{S} \frac{1}{x}$$

Eq. 9

## Third null hypothesis: sampling partitions, $H_{null}^{part}$

### Probability of a given SAD

We can choose a direct sampling of partitions, without considering the underlying distribution of compositions or surjections. Sampling partitions results in an unbiased sample of unique SAD shapes and represents the observable variation in the form of the pattern without considering how that pattern relates to the abundances of the individual (labelled) species. This is the sampling that results on more uneven SADs, closer to most empirical SADs (Locey & White 2013).

When sampling partitions according to the distribution of partitions only, the probability of a given partition $SAD = \{n_1, n_2, n_3, ..., n_S\}$, $n_1 \geq n_2 \geq n_3 \geq ... \geq n_S \geq 1, \sum_{i=1}^{S} n_i = N$, is:

$$P(partition\ SAD|N,S) = \frac{number\ of\ partitions\ per\ partition}{number\ of\ partitions} = \frac{1}{p(N,S)}$$

Eq. 10

, where $p(N,S)$ is the number of partitions of $N$ into exactly $S$ parts.

There is no closed expression for the value of $p(N,S)$. It is the coefficient of $z^S x^N$ in $\prod_{a=1}^{N-S+1}(1 - zx^a)^{-1}$ (Gupta 1970). It follows the recurrence relation $p(N,S) = p(N-1, S-1) + p(N-S, S)$ (Gupta 1970). An efficient calculation of $p(N,S)$ has been implemented in function *NrParts* in the R package *rpartitions* (Locey & McGlinn 2012). In any case, the value of $p(N,S)$ tends to $p(N,S) \approx \frac{N^{S-1}}{S((S-1)!)^2}$ as $N \to \infty$ (Knessl & Keller 1990). This asymptotic expression is not valid when $S$ is too large or, more specifically, if $S$ grows as a function of $\sqrt{N}$ (see Knessl & Keller 1990 for alternatives). However, this will not be the case in most ecological scenarios: once one has sampled many individuals, the increase in diversity per new individual is typically small and slowly tending to zero.

*What are the typical SADs expected under $H_{null}^{part}$? And the most likely SAD?*

Under $H_{null}^{part}$, unlike in the previous cases, all partitions are equally likely. Therefore, there is not such a thing as 'the most likely partition' in the absence of any biological or ecological mechanism, given $N$ and $S$. However, some *general shapes* can be more common than others. In fact, most partitions tend to be hollow curves very similar to the log-normal distribution and empirical SADs (Locey & White 2013). Unfortunately, sampling partitions directly (regardless of the distribution of compositions) is much more difficult than sampling surjections or compositions. In fact, computational difficulties have been identified as a major obstacle to the application of the feasible set for the study of SADs (Locey & McGlinn 2014). These authors describe relatively fast algorithms for the random sampling of partition SADs. The code needed to run these algorithms has been implemented in the R package *rpartitions* (Locey & McGlinn 2012) and is not described here.

### Comparing the null expectations with observations and alternative models

Once the expected SAD is defined under the selected null hypothesis, we can compare it with the observed partition SAD using the squared Pearson's $r$ ($r^2$) between the observed and expected SADs. This statistic, $r^2$, can be interpreted as the proportion of the variation in species abundances that can be explained by $N$ and $S$ alone (Marks & Muller-Landau 2007; Locey & White 2013).

After taking many random draws from the feasible set, we can calculate a standardized effect size, e.g. a $z$-score, to estimate the magnitude of the deviation from the expectation by the null hypothesis. This considers not only the expected shape (the center of the feasible set) but also the variability in the expected shape (the density of the feasible set around its center):

$$z = \frac{observed\ SAD - mean(null\ simulations)}{sd(null\ simulations)}$$

Eq. 11

Note that one would have $S$ values of $z$: $z_1, z_2, ..., z_S$. Thus, from the null model one can distinguish if a given observed SAD deviates from the expectations in the tail of rare species, or in the most common species, or in particular regions of the rank-abundance distribution. The distribution of $z$ values itself could be modelled and interpreted ecologically.

The expectations by the null hypotheses described here can be included in a broader comparison among SAD models. $P(partition\ SAD|S,N)$ is the likelihood of observing the data, given the null hypothesis. It can be compared with the probability of that same partition SAD given any other model and parameters (i.e., the likelihood of any particular

SAD model fitted to the data). One way is by using the corrected Akaike information criterion (AIC$_C$). The calculation of the AIC$_C$ for many SAD models is possible, and should be seen as best practice for species abundance distributions (Connolly *et al.* 2014; Matthews & Whittaker 2014; Baldridge *et al.* 2016). In general, AIC$_C$ is calculated based on the likelihood of a model ($L$), the number of observations ($n$) and the number of free parameters in the model ($k$): $AIC_C = 2k - 2\log(L) + \frac{2k(k+1)}{n-k-1}$; where $n$ equals $S$, not $N$: SADs are curves fitted to $S$ points. All the null hypotheses related to the feasible set approach have no free parameters, thus $k = 0$ and $AIC_{C\,null} = -2\log\bigl(P(partition\,SAD|H_{null}, N, S)\bigr)$; where $P(partition\,SAD|H_{null}, S, N)$ is a different value depending on the version of the null hypothesis tested. The AIC$_C$ values obtained for the null and alternative models can be compared as usual (Burnham & Anderson 2002; Matthews & Whittaker 2014; Baldridge *et al.* 2016) to determine whether a given alternative model deviates sufficiently from the expectation by the null hypothesis chosen.

**Future research and some variations of the feasible set approach**

*Hard vs. soft constraints*

I have described feasible sets based on hard constraints. It would be possible to allow soft constraints instead. We may generate expectations that vary in the specific number of individuals, but the average of all the simulations would be $N$. In this case, $N$ would be a weak constraint (Haegeman & Etienne 2010). This links to the (very plausible) idea that the specific $N$ in the sample is not a strict property of the system, but a specific instance of a more general phenomenon (the density of individuals in the system and its variability). The interpretation of soft constraints can be very ecological, on the other hand. For example, we can consider that the studied region is composed by different assemblages (or habitats), each assemblage $x$ characterized by a different number of species, $S_x \leq S_{regional}$, so the region would be characterized by a given distribution of the probability $P(S_x)$, $\int P(S_x) = 1$. The expected SAD in the sample would result from choosing $S_x$ according to $P(S_x)$ to generate the feasible set. Here, $P(S_x)$ provides soft constraints with (potentially) ecological interpretation. Another approach using soft constraints is to generate feasible sets at the regional level, either using hard or soft constraints, and then adding the sampling process to make predictions on the sample SAD. For example, the feasible set constrained by $N_{region}$ ($N_{region} \to \infty$) and an estimation of $S_{regional}$ could generate the expected regional SAD. Then, one could calculate the probability of the observed SAD in the sample, given the expected regional SAD, using the negative binomial distribution, or other distributions (Green & Plotkin 2007; Connolly, Hughes & Bellwood 2017). In this case, sample $S$ would be weakly constrained through its upper bound, $S_{regional}$. Even if an analytical solution is difficult to derive, the algorithms for numerical simulations already exist, and this can be repeated may times to generate an average SAD.

*The composition matrix*

Solving the feasible set for composition matrixes is among the most promising challenges in constraint-based ecology, from my point of view. Composition matrixes ($P \times S$ matrixes containing the abundances of $S$ species into $P$ plots) are the combination of $P$ weak integer compositions (zero's allowed) in the sample. If the probability of observing a given matrix $P \times S$ can be calculated based on $N$ and $S$, then the feasible set approach can be applied to a number of ecological relationships involving local aggregation, species-abundance distributions, and species associations, in a unified framework (Connolly *et al.* 2017). Generating many compositions independently (one per sample) and then pasting them together cannot generate a useful expected $P \times S$. We would be simulating a situation where regional processes do not constraint at all what one can see in the sample, and we would not get anything close to the regional SAD. Therefore, the $P \times S$ matrix should be constrained, at least, by the column sums (the SAD when all the samples are pooled together). There is a rich yet rather impenetrable body of mathematical literature related to the number (~probability) of integer matrixes with prescribed marginal sums, i.e., with a given pooled

SAD and a given distribution of total abundances within the plots (e.g. Barvinok 2010, 2012; Barvinok & Hartigan 2012). It would be a difficult but rewarding task to distill that message into ecology. On the other hand, there are more-or-less efficient algorithms to sample uniformly presence/absence matrixes with prescribed marginal sums from the feasible set (e.g. Chakraborty *et al.* 2003; Chen *et al.* 2005; Bezáková, Bhatnagar & Vigoda 2007). Maybe these approaches can be expanded into more general integer matrices, with values >1 allowed.

I ignore if the asymptotic behavior as $N \to \infty$, with no zero's in the matrix, results more tractable, but it is worth to explore. At the population level, we can think of the composition matrix as the combination of $P$ natural compositions, with strictly positive and continuous abundances in each cell, constrained by the column sums (i.e, the asymptotic regional SAD, which would be in turn constrained by $S_{regional}$). If the typical shape of such a matrix can be determined, or the set of feasible matrixes can be sampled uniformly, one can generate null integer composition matrixes by sampling from the continuous matrix according to the observed total number of individuals at each plot. This would generate null predictions about patterns of species co-occurrences, patterns of species aggregation, alpha- and beta-diversity, etc., strictly based on the regional diversity and the characteristics of the sample (number of samples and number of individuals per sample).

*Dynamic feasible sets*

Another combinatorial approach to the feasible set would be to add more realistic constraints in the upper (and possibly lower) bounds of species abundances in the possible integer compositions underlying the SAD. For example, one could calculate the number of sequences $(n_1, n_2, n_3)$ satisfying $n_1 + n_2 + n_3 = 30$, subject to the constraints $0 \leq n_1 \leq 10$, $0 \leq n_2 \leq 20$ and $5 \leq n_3 \leq 30$, if we have some prior knowledge about the regional distribution of those three species. Similar constraints would exclude the possibility, when building the feasible sets, that the rarest species in the region accumulates ~$N$ individuals in the sample. (If that happens, that assemblage is probably not the object of study). Or, if we are sampling a pineland, similar constraints would exclude from the feasible set the compositions with less than $N/2$ pine individuals, for example. Yes, pine-dominated forests could be the result of chance, if Nature is blindly sampling compositions from the pool of possible compositions. But, if we are intentionally sampling >90% pine stands by excluding other forest types, we cannot get a non-pine stand in our sample "just by chance", and we should not count that as one of the possible outcomes of the sampling process.

We could, in theory, add constraints sequentially until the center of the feasible set reaches a high match with the reality (e.g. $r^2 = 0.95$). The goal of a dynamic constraining algorithm would be to find the weakest constraint required on the underlying compositions to reach a given fit between the expected average partition and the observed partition. I am not aware of anyone exploring the gradient between null models and not-so-null versions of those same models. Increasing constraints in purely combinatorial null models could be a way of describing such gradient with a very high resolution. The amount of constraints required would quantify the (minimum) weight of regional ecological processes required to explain the local sample SAD in the absence of any local assembly mechanism.

**Conclusion**

Locey & White (2013) provided a simple and intuitive description of what should be expected in the absence of any biological or ecological mechanism. A major merit of their work was to present the concept of the feasible set to a broad audience of ecologists,. Here I present a more complete summary of the available tools, for three different definitions of the feasible set. In general, it is important to consider the SAD shape relative to the feasible set, rather than the SAD shape *per se* (Locey & White 2013). There exist explicit mathematical descriptions of the distribution of partitions and accessible algorithmic implementations to simulate expectations numerically (see Table 1 for a concise summary). The integration with the typical framework of inference in ecology (goodness-of-fit, null hypothesis testing, model comparison, null modelling) is straightforward. It should not be difficult to consider the null hypotheses derived from the feasible set when testing for any mechanism underlying the SAD.

**Acknowledgements ·** This work was inspired by comments of Alberto Jiménez-Valverde and one anonymous referee during the review process of Arellano *et al.* (2017). Kenneth Locey provided clarifications about Locey & White (2013), highly stimulating discussion, and very useful comments and editions to this manuscript. I am extremely grateful for his contributions. Mª Natalia Umaña and Luis Cayuela provided useful comments on different drafts of this work. I wrote this note as a multi-year side project while being supported by the Comunidad Autónoma de Madrid (Spain) and the Next Generation Ecosystem Experiments-Tropics, funded by the US Department of Energy, Office of Science, Office of Biological and Environmental Research.

# References

Arellano, G., Umaña, M.N., Macía, M.J., Loza, M.I., Fuentes, A., Cala, V. & Jørgensen, P.M. (2017) The role of niche overlap , environmental heterogeneity , landscape roughness and productivity in shaping species abundance distributions along the Amazon – Andes gradient. *Global Ecology and Biogeography*, **26**, 191–202.

Baldridge, E., Harris, D.J., Xiao, X. & White, E.P. (2016) An extensive comparison of species-abundance distribution models. *PeerJ*, **4**, e2823.

Barvinok, A. (2010) On the number of matrices and a random matrix with prescribed row and column sums and 0-1 entries. *Advances in Mathematics*, **224**, 316–339.

Barvinok, A. (2012) Matrices with prescribed row and column sums. *Linear Algebra and Its Applications*, **436**, 820–844.

Barvinok, A. & Hartigan, J.A. (2012) An asymptotic formula for the number of non-negative integer matrices with prescribed row and column sums. *Transactions of the American Mathematical Society*, **364**, 4323–4368.

Bezáková, I., Bhatnagar, N. & Vigoda, E. (2007) Sampling binary contingency tables with a greedy start. *Random Structures and Algorithms*, **30**, 168–205.

Burnham, K.P. & Anderson, D.R. (2002) *Model Selection and Multimodel Inference*. Springer-Verlag, New York, U.S.A.

De Cáceres, M. & Legendre, P. (2008) Beals smoothing revisited. *Oecologia*, **156**, 657–669.

Chakraborty, A., Chen, Y., Diaconis, P., Holmes, S. & Liu, J.S. (2003) *Sequential Monte Carlo Methods for Statistical Analysis of Tables*. Department of Statistics, Stanford University.

Chen, Y., Diaconis, P., Holmes, S.P. & Liu, J.S. (2005) Sequential Monte Carlo methods for statistical analysis of tables. *Journal of the American Statistical Association*, **100**, 109–120.

Conlisk, J., Conlisk, E., Kassim, A.R., Billick, I. & Harte, J. (2012) The shape of a species' spatial abundance distribution. *Global Ecology and Biogeography*, **21**, 1167–1178.

Connolly, S.R., Hughes, T.P. & Bellwood, D.R. (2017) A unified model explains commonness and rarity on coral reefs. *Ecology letters*.

Connolly, S.R., MacNeil, M.A., Caley, M.J., Knowlton, N., Cripps, E., Hisano, M., Thibaut, L.M., Bhattacharya, B.D., Benedetti-Cecchi, L., Brainard, R.E., Brandt, A., Bulleri, F., Ellingsen, K.E., Kaiser, S., Kröncke, I., Linse, K., Maggi, E., O'Hara, T.D., Plaisance, L., Poore, G.C.B., Sarkar, S.K., Satpathy, K.K., Schückel, U., Williams, A. & Wilson, R.S. (2014) Commonness and rarity in the marine biosphere. *Proceedings of the National Academy of Sciences of the United States of America*, **111**, 8524–9.

Etienne, R.S. & Olff, H. (2005) Confronting different models of community structure to species-abundance data: a Bayesian model comparison. *Ecology letters*, **8**, 493–504.

Green, J.J.L. & Plotkin, J.B.J. (2007) A statistical theory for sampling species abundances. *Ecology Letters*, **10**, 1037–45.

Gupta, H. (1970) Partitions - A Survey. *Journal of Research of the National Bureau of Standards - B. Mathematical Sciences*, **74**, 1–29.

Haegeman, B. & Etienne, R.S. (2010) Entropy Maximization and the Spatial Distribution of Species. *The American Naturalist*, **175**, E74–E90.

Haegeman, B. & Loreau, M. (2008) Limitations of entropy maximization in ecology. *Oikos*, **117**, 1700–1710.

Harte, J. (2003) Ecology: tail of death and resurrection. *Nature*, **424**, 1006–1007.

Knessl, C. & Keller, J.B. (1990) Partition Asymptotics From Recursion Equations. *SIAM Journal on Applied Mathematics*, **50**, 323–338.

Locey, K. & McGlinn, D. (2012) rpartitions: code for integer partitioning.

Locey, K. & McGlinn, D. (2014) Efficient algorithms for sampling feasible sets of abundance distributions. *PeerJ PrePrints*, 2:e78v2.

Locey, K.J. & White, E.P. (2013) How species richness and total abundance constrain the distribution of abundance. *Ecology Letters*, **16**, 1177–1185.

MacArthur, R.H. (1957) On the relative abundance of bird species. *Proceedings of the National Academy of Science of the United States of America*, **43**, 293–295.

Marks, C.O. & Muller-Landau, H.C. (2007) Comment on "From Plant Traits to Plant Communities: A Statistical Mechanistic Approach to Biodiversity."

*Science*, **316**, 1425c–1425c.

Matthews, T. & Whittaker, R. (2014) Fitting and comparing competing models of the species abundance distribution: assessment and prospect. *Frontiers of Biogeography*, **6**, 67–82.

McGill, B.J. (2003a) Does Mother Nature really prefer rare species or are log-left-skewed SADs a sampling artefact? *Ecology Letters*, **6**, 766–773.

McGill, B.J. (2003b) A test of the unified neutral theory of biodiversity. *Nature*, **422**, 881–885.

McGill, B.J. (2010) Towards a unification of unified theories of biodiversity. *Ecology Letters*, **13**, 627–642.

McGill, B.J. (2011) Species abundance distributions. *Biological diversity: frontiers in measurement and assessment* (eds A.E. Magurran & B.J. McGill), pp. 105–122. Oxford University Press, Oxford, UK.

McGill, B.J., Etienne, R.S., Gray, J.S., Alonso, D., Anderson, M.J., Benecha, H.K., Dornelas, M., Enquist, B.J., Green, J.L., He, F., Hurlbert, A.H., Magurran, A.E., Marquet, P.A., Maurer, B. a, Ostling, A., Soykan, C.U., Ugland, K.I. & White, E.P. (2007) Species abundance distributions: moving beyond single prediction theories to integration within an ecological framework. *Ecology Letters*, **10**, 995–1015.

Nee, S. & Stone, G. (2003) The end of the begining for neutral theory. *Trends in Ecology and Evolution*, **18**, 433–434.

Nekola, J.C. & Brown, J.H. (2007) The wealth of species: ecological communities, complex systems and the legacy of Frank Preston. *Ecology Letters*, **10**, 188–196.

Pueyo, S., He, F., Zillio, T. & Salvador Pueyo, F.H.T.Z. (2007) The maximum entropy formalism and the idiosyncratic theory of biodiversity. *Ecology Letters*, **10**, 1017–1028.

Šizling, A., Storch, D., Sizlingova, E., Reif, J. & Gaston, K.J. (2009) Species abundance distribution results from a spatial analogy of central limit theorem. *Proceedings of the National Academy of Sciences*, **106**, 6691–6695.

Warren, R.J., Skelly, D.K., Schmitz, O.J. & Bradford, M. a. (2011) Universal ecological patterns in college basketball communities. *PloS One*, **6**, e17342.

White, E.P., Thibault, K.M. & Xiao, X. (2012) Characterizing species abundance distributions across taxa and ecosystems using a simple maximum entropy model. *Ecology*, **93**, 1772–1778.

Wolters, M.A. (2015) A Genetic Algorithm for Selection of Fixed-Size Subsets with Application to Design Problems. *Journal of Statistical Software, Code Snippets*, **68**, 1–18.

**Table 1**. Summary of the results presented in this study, according to three main null hypotheses: SADs as a result of Nature sampling randomly surjections, compositions, or partitions, respectively. These three hypotheses describe expectations in the absence on any biological or statistical mechanism. The "probability of a given SAD" refers to the probability of a given partition, since "SAD shape" always refer to partitions in the ecological literature (i.e. two communities can have the same SAD shape even if they have different species or the same species have different abundances). In some cases, the mot likely SAD can be calculated using numerical optimization or using aymptotic solutions. In all cases, typical shapes can be estimated by sampling repeatedly from the feasible set.

| | What is Nature sampling blindly? | Probability of a given SAD | How to sample from the feasible set | The most likely SAD |
|---|---|---|---|---|
| $H_{null}^{surj}$ | Surjections: sequences of $N$ individuals with random taxonomic identities. E.g. (A, B, B, A, B) | Eq. 4 | Easy; see Appendix 1 | That maximizing Eq. 6. Optimization of $S$ parameters required, see Appendix 3. As $N \to \infty$, it approaches perfectly even abundances, with all species having $N/S$ individuals |
| $H_{null}^{comp}$ | Compositions: sequences of $S$ random abundances with respect to which species has which abundance (i.e., order matters). E.g. $\{A = 2, B = 3\}$ | Eq. 8 (Etienne & Olff 2005) | Easy because of the equivalence with the broken stick model; see Appendix 2 | That minimizing $\sum_{n=1}^{N}(\Phi_n \log \Phi_n - \Phi_n)$. Optimization of $S$ parameters required, see Appendix 3. As $N \to \infty$, the expected by the broken stick model, Eq. 9 (MacArthur 1957) |
| $H_{null}^{part}$ | Partitions: sequences of $S$ random abundances without respect to which abundance belongs to which species. E.g. $\{3, 2\}$ | It is $1/p(N,S)$. The exact value of $p(N,S)$ can be calculated recursively using the function *NrParts* in the *rpartitions* package (Locey & McGlinn 2012). The asymptotic value of $p(N,S)$ is $\frac{N^{S-1}}{S((S-1)!)^2}$ (Knessl & Keller 1990) | Non-trivial; see *rpartitions* package (Locey & McGlinn 2012) | All SADs are equally likely |

**Supplementary material**

**Appendix 2**. R code to generate a partition resulting from a random surjection.

```
random_surjection  <- c(1:S, sample(1:S, size = N – S, replace = TRUE))
composition <- table(random_surjection)
partition    <- as.numeric(sort(composition, decreasing = TRUE))
```

**Appendix 2**. R code to generate a partition resulting from a random composition.

```
# Integer case:
random_breaks <- c(0, sample(1:(N-1), size = S-1), N) + 0.5
random_composition <- table(cut(1:N, breaks = random_breaks))
partition <- as.numeric(sort(random_composition, decreasing = TRUE))

# Continuous case, as N tends to infinite:
random_breaks <- sort(c(0, runif(min = 0, max = 1, n = S - 1), 1)
random_composition <- random_breaks[2:(S+1)] - random_breaks[1:S]
partition <- as.numeric(sort(random_composition, decreasing = TRUE)))
```

**Appendix 3**. Estimation of the most likely SAD according to $H_{null}^{surj}$ or $H_{null}^{comp}$.

The most likely SAD according to $H_{null}^{surj}$ or $H_{null}^{comp}$ can be estimated by optimizing numerically $S$ parameters. The probabilities involved require calculating huge numbers, which makes difficult to optimize directly on the exact expressions. The R code below maximizes $\sum_{i=1}^{S-1}\left(\left(\sum_{j=i}^{S} n_j\right)\log\left(\sum_{j=i}^{S} n_j\right)\right) - \sum_{i=1}^{S-1}(n_i \log n_i) - \sum_{i=1}^{S-1}\left(\left(\sum_{j=i}^{S} n_j - n_i\right)\log\left(\sum_{j=i}^{S} n_j - n_i\right)\right) - \sum_{n=1}^{N}(\Phi_n \log \Phi_n - \Phi_n)$ in the case of $H_{null}^{surj}$, or minimizes $\sum_{n=1}^{N}(\Phi_n \log \Phi_n - \Phi_n)$ in the case of $H_{null}^{comp}$; $n_i$ being the abundance of the species $i$ and $\Phi_n$ the number of species with $n$ individuals. As $N \to \infty$ the expected SAD under $H_{null}^{surj}$ are perfectly even abundances, all species with $N/S$ individuals. As $N \to \infty$ the expected SAD under $H_{null}^{comp}$ is the expected by the continuous broken stick model (MacArthur 1957). These asymptotic expectations are used to initialize the numerical optimization in each case.

There are five functions in this appendix:

- `adj`: an auxiliary adjustment routine, taking continuous abundances and returning integer abundances. This is to facilitate the optimization based on continuous (instead of integer) parameters.
- `f_surjections`: a function to minimize under $H_{null}^{surj}$.
- `f_compositions`: a function to minimize under $H_{null}^{comp}$.
- `mlSAD_surjections`: a wrapper for the base R *optim* function. It performs the optimization, and returns the most likely integer SAD under $H_{null}^{surj}$. The optimization is not done directly on species abundances, but on $S - 1$ breaking points to ensure that the species-level abundances sum up to the total abundance.
- `mlSAD_compositions`: similar to mlSAD_surjections, it returns the most likely integer SAD under $H_{null}^{comp}$.

The code below is presented only with didactic purposes, as an example. There are many different alternatives to perform numerical optimization. Better, more complex code should probably involve integer optimization of $S-1$ integer parameters with repetitions not allowed. The *kofnGA* R package implements a genetic algorithm that may be useful for that purpose (Wolters M.A. 2015. A Genetic Algorithm for Selection of Fixed-Size Subsets with Application to Design Problems. *Journal of Statistical Software* 68, 1–18).

```
### A general adjustment routine, from continuous abundances to positive
integers:
adj <- function(n, N, S)
{
  if(length(n) != S) stop("n must have S length.")
  n <- sort(as.integer(round(N * n / sum(n))), decreasing = TRUE)

  while(sum(n == 0) > 0 | sum(n) != N)
  {
    n[n == 0] <- 1
    d = N - sum(n)

    if(d != 0)
    {
      ED <- density(n, from = 1, to = max(n), kernel = "epanechnikov")
      if(d > 0) i = which.min(abs(n - ED$x[which.max(ED$y)]))
      if(d < 0)
      {
        j <- which(n > 1)
        k = which.min(abs(n[j] - ED$x[which.max(ED$y)]))
        i = j[k]
      }
      n[i] <- n[i] + ifelse(d > 0, +1, -1)
      n <- sort(n, decreasing = TRUE)
    }
  }
  n
}

### Two functions to minimize:
f_surjections <- function(breaks, N)
{
  # Generate a vector of abundances
  S = length(breaks) + 1
  breaks <- sort(c(0, breaks, 1))
  n0 <- breaks[2:(S+1)] - breaks[1:S]
  n <- adj(n0, S = S, N = N)

  # Implement Equation 6, a value to maximize
  Eq_6 <- function(n)
  {
    # Summands involving sums along S-1
    S = length(n)
    s1 <- sapply(1:(S-1), function(i) sum(n[i:S]))
    s2 <- s1 - n[1:(S-1)]
    a1 = sum(s1 * log(s1))
    a2 = sum(n[1:(S-1)] * log(n[1:(S-1)]))
    a3 = sum(s2 * log(s2))

    # Summand involving sum along N
    t <- table(n)
    a4 = sum(t * log(t) - t)
```

```r
    # Return:
    a1 - a2 - a3 - a4
  }

  # Return the value to minimize by "optim"
  return(-Eq_6(n))
}

f_compositions <- function(breaks, N)
{

  # Generate a vector of abundances
  S = length(breaks) + 1
  breaks <- sort(c(0, breaks, 1))
  n0 <- breaks[2:(S+1)] - breaks[1:S]
  n <- adj(n0, S = S, N = N)

  # Implementation to a proxy for Equation 8,
  # a value to minimize:
  proxy_to_Eq_8 <- function(n)
  {
    t <- table(n)
    sum(t * log(t) - t)
  }

  # Return the value to minimize by "optim"
  proxy_to_Eq_8(n)
}

### Two wrapers for the optimization that return the
### most likely SAD according to both null hypotheses:
mlSAD_surjections <- function(N, S, …)
{
  # Optimization initialized with even abundances:
  parzero <- seq(0, 1, length.out = S - 1)
  o <- optim(par = parzero, fn = f_surjections, N = N, lower = 0, upper = 1, method = "L-BFGS-B", …)

  # The most likely SAD:
  breaks <- sort(c(0, o$par, 1))
  mlSAD <- adj(n = breaks[2:(S+1)] - breaks[1:S], S = S, N = N)
  return(mlSAD)
}

mlSAD_compositions <- function(N, S, …)
{
  # Optimization initialized with the broken stick model
  # (this is typically the best possible solution):
  parzero <- cumsum(sapply(1:(S-1), function(i) 1/S*sum(1/(i:S))))
  o <- optim(par = parzero, fn = f_compositions, N = N, lower = 0, upper = 1, method = "L-BFGS-B", …)

  # The most likely SAD:
  breaks <- sort(c(0, o$par, 1))
  mlSAD <- adj(n = breaks[2:(S+1)] - breaks[1:S], S = S, N = N)
  return(mlSAD)
}
```